\documentclass[a4paper,11pt]{article}
\pdfoutput=1 

\usepackage{jheppub} 
             
\usepackage{amsmath,amssymb,epsfig,color}                     
\usepackage{lipsum}
\usepackage{latexsym}
\usepackage{url}
\usepackage{amsthm}
\usepackage{acronym}
\usepackage{graphicx}
\usepackage{hyperref}
\graphicspath{ {} }                     

\usepackage[T1]{fontenc}

\def\exp{{\rm exp}\,}

\title{\boldmath Exotic N dependent free energy of black brane solutions from 3D MSYM }

\author[a]{Ankit Vikrant}

\affiliation[a]{BITS Pilani Goa Campus,Goa, India}

\emailAdd{ankitvikrant74@gmail.com}

\abstract{Recently it was found in \cite{Jafferis:2015ppr1} that dyonically-gauged N=8 supergravity arises as a consistent truncation of massive type IIA supergravity when the gauge group is ISO(7). In particular, they found a critical point of the supergravity that uplifted to the first explicit N = 2 $AdS_4$ massive IIA background. Its free energy was also calculated.
Though no black brane solutions for this theory have been constructed yet, we nevertheless expect the free energy of the black brane solutions to scale with the same N dependence as the solution obtained in \cite{Jafferis:2015ppr1}. In this note, we will reproduce the exotic N dependence of the free energy of such black brane solutions from field theory at finite temperature using the Smilga-Wiseman approach. The free energy expression thus obtained also tells us about the temperature dependence of free energy of such black brane solutions in a given low temperature regime.}

\begin{document} 
\maketitle
\flushbottom

\section{Introduction}
\label{sec:intro}

In a recent paper \cite{Jafferis:2015ppr1}, it was found that ISO(7)-dyonically-gauged $\mathcal{N} = 8$ supergravity has a higher dimensional origin in massive type IIA supergravity on $S^6$. They prescribe a particular consistent embedding that allows one to uplift any four dimensional $\mathcal{N}=8$ ISO(7) supergravity solution to massive type IIA. It also preserves supersymmetry in the process. They also find a particular critical point and uplift it to obtain the first explicit, analytic $N$=2 $AdS_4$ solution in massive type IIA background. Further, it's gravitational free energy was also calculated for comparison with the free energy of the candidate dual $\mathcal{N}=2$ Chern-Simons-matter theory that they describe. Another solution was constructed in \cite{Rong:2015ppr1} for the $\mathcal{N} = 3$ case. They too found a dual for the solution they constructed, and entrenched the argument for duality by comparing the free energies of bulk and boundary \cite{Rong:2015ppr2}. Quite clearly, here too the free energy had the same N scaling as for the solution constructed in \cite{Jafferis:2015ppr1}.

One would expect a similar N dependence in the free energy expressions obtained from black-hole solutions of this particular supergravity but such solutions have not been constructed yet. In this brief note, we will try to obtain the form of the free energy of the possible black-brane solutions from field theory using Smilga-Wiseman method. The free energy expression thus obtained has a temperature dependence, and we predict that such dependence would also be seen in free energy of black-brane solutions obtained from supergravity.

\section{Free energy calculation from field theory via Smilga-Wiseman method}
\label{sec:Maincalc}

In \cite{Jafferis:2015ppr1}, the authors considered a dual field theory which was an $\mathcal{N}$ = 2 supersymmetric SU(N) SYM gauge theory with a Chern-Simons term of level k. This theory is a deformation of the $\mathcal{N}$ = 8 maximally supersymmetric Yang-Mills with gauge group SU(N). The $AdS_4$ solutions constructed by them in the bulk, naturally arise as near horizon geometries of D2 branes in smooth backgrounds of massive type IIA owing to the topology of the internal manifold. In flat space, the worldvolume theory of N D2 branes in massless IIA is the maximal SU(N) SYM in three dimensions. We'll be working with a Euclidealized action for this field theory at finite temperature to calculate the free energy expression. Finally, we'll see that our expression holds for a certain temperature limit which is analogous to having a low curvature condition in the dual supergravity. \\

At finite temperature, the Euclideanized action for 3 dimensional maximally supersymmetric SU(N) Yang-Mills would be: \\

\begin{equation} \label{eu_eqn12}
\begin{split}
S_{\text{D}2} = \frac{N}{\lambda}\int d\tau d^2x Tr\Big[ \frac{1}{4} \mathcal{F}_{\mu\nu}^2 + \frac{1}{2}(D_\mu\Phi^I)^2 + \frac{i}{2}\bar\Psi\Gamma^\mu D_\mu\Psi - \\
\frac{1}{4}[\Phi^I,\Phi^J]^2 - \frac{i}{2}\bar\Psi\Gamma^I[\Phi^I,\Psi] \Big]
\end{split}
\end{equation}

where Euclidean time $\tau=ix^0$ is periodic with a period of $\beta=1/T$. Also,$\mu=0,\ldots,2$ and $I=1,\ldots,7$. 
The configurations of classical vacua that we consider here are: \\
\begin{align} \label{eu_eqn13}
A_{\mu,ab}&=a_{\mu,a}\delta_{ab},                &           \Phi^I_{ab}&=\phi^I_a\delta_{ab},     &       
\Psi_{ab}=0,
\end{align}

where $a_{\mu,a}$ and $\phi_a^I$ are real constants.% 
We assume that only scalar moduli are relevant. These scalar moduli, $\phi_a^I$ represent the positions of the D2-branes in the transverse directions. Also, $a,b=1,\ldots, N$ are the indices of the adjoint hermitian matrices. The subsequent calculation draws from the calculation for Dp-branes (p=2 here) in \cite{Morita:2013ppr1} , 

We assume that at low temperature, the dominant configurations of the D2-branes satisfy 
\begin{align} \label{eu_eqn14}
 \beta |\phi_a-\phi_b| \gg 1\,,
 \end{align}
 
where $|\phi_a-\phi_b|:=\sqrt{\sum_I(\phi_a^I-\phi_b^I)^2}$.
In this case, the D2-branes are separated by large distances and the off-diagonal components possess large mass $\sim |\phi_a-\phi_b|$, so they can be integrated out. Then, the low temperature dynamics would be governed by effective theory of the scalar moduli $\phi^I_a$.

From the classical action (\ref{eu_eqn12}), we obtain the classical term
\begin{align} \label{eu_eqn15}
S_{\text{D}2}^{\text{cl}} = \frac{N}{\lambda} \int d \tau d^2x \sum_{a} \left( \frac{1}{2} \partial^\mu \phi^I_a \partial_\mu \phi^I_a \right).
\end{align}

 The integrals of the massive fields induce the quantum corrections .
We are only considering the leading one-loop correction.
We look at temperature independent corrections, which don't have an explicit temperature dependence, and temperature dependent terms.
The one-loop effective action at zero temperature is given by
\begin{equation} \label{eu_eqn16}
\begin{split}
S_{\text{D}2,T=0}^{\text{1-loop}} = -\int d \tau d^2x \sum_{a<b} \frac{\Gamma\left(\frac{5}{2}\right)}{(4\pi)^\frac{3}{2}} \Big[ 2\frac{\{\partial_\mu (\phi_a^I -\phi_b^I) \partial_\nu (\phi_a^I -\phi_b^I)\}^2}{|\phi_a -\phi_b|^{5}} \\
-\frac{ \{ \partial_\mu  (\phi_a^I -\phi_b^I) \partial^\mu (\phi_a^I -\phi_b^I)\}^2 }{|\phi_a -\phi_b|^{5}}\Big] + \ldots .
\end{split}
\end{equation}

The quantum correction starts from the quartic derivative terms $(\partial \phi)^4$ because supersymmetry ensures that quadratic terms $(\partial \phi)^2$ don't receive any corrections. We haven't included the subdominant terms, which are suppressed when the distances between the branes are large and $ | (\partial \phi)^2  /\phi^4| \ll 1$ is satisfied. \\

The temperature dependent terms are proportional to $\exp(-\beta|\phi_a -\phi_b|)$, and hence suppressed under the condition (\ref{eu_eqn14}). \\

Now there are 2 important assumptions for $\phi$ and derivative terms. The first one is as follows:
\begin{equation} \label{eu_eqn17}
\phi^I_a  \sim \phi^I_a-\phi^I_b \sim \phi.
\end{equation}

Roughly speaking, this means that all the branes are uniformly distributed, i.e. , all distances are of the same order. \\

To determine the scale $\phi$, we also assume that the derivative terms satisfy
\begin{equation} \label{eu_eqn18}
  \partial \phi^I_a  \sim \partial (\phi^I_a-\phi^I_b) \sim \frac{1}{\beta} \phi .
\end{equation}

This important assumption implies that the temperature dominates the moduli dynamics rather than the 't Hooft coupling $\lambda$. \\

By using the assumptions (\ref{eu_eqn17}) and (\ref{eu_eqn18}), we compute $S_{\text{D}2}^{\text{cl}}$ first.

\begin{align} \label{eu_eqn19}
  S_{\text{D}2}^{\text{cl}} = \frac{N}{\lambda} \int d \tau d^2x \sum_{a} \left( \frac{1}{2}  {\frac{\phi^2 }{\beta^2}} \right) .
\end{align}

By using $\int d\tau \sim \beta$ and $\sum_a \sim N$, we are left with: 

\begin{equation} \label{eu_eqn20}
S_{\text{D}2}^{\text{cl}}  \sim  \int d^2x \frac{N^2}{\beta \lambda} \phi^2
\end{equation}

Let's compute $S_{\text{D}2,T=0}^{\text{1-loop}}$ now using (\ref{eu_eqn17}) and (\ref{eu_eqn18}) again:

\begin{equation} \label{eu_eqn21}
\begin{split}
S_{\text{D}2,T=0}^{\text{1-loop}} = -\int d \tau d^2x \sum_{a<b} \frac{\Gamma\left(\frac{5}{2}\right)}{(4\pi)^\frac{3}{2}}\Big[ 2\frac{\{\partial (\phi) \partial (\phi)\}^2}{\phi^{5}} \\
-\frac{ \{ \partial  (\phi) \partial (\phi)\}^2 }{\phi^{5}}\Big] + \ldots .
\end{split}
\end{equation}

\[ = -\int d \tau d^2x \sum_{a<b} \frac{\Gamma\left(\frac{5}{2}\right)}{(4\pi)^\frac{3}{2}}\Big[ 2\frac{{\frac{\phi^2}{\beta^2}}^2}{\phi^{5}} - \frac{ { \frac{\phi^2}{\beta^2} }^2 }{\phi^{5}}\Big] + \ldots \]

We can further use $\int d\tau \sim \beta$ and $\sum_{a<b} \sim N^2$ : \\

\[ S_{\text{D}2,T=0}^{\text{1-loop}} = -\int d^2x \beta N^2  \frac{\Gamma\left(\frac{5}{2}\right)}{(4\pi)^\frac{3}{2}}\Big[ 2 {\frac{\phi^{-1}}{\beta^4}} - \frac{\phi^{-1}}{\beta^4}\Big] + \ldots \]

Finally, we are left with: 

\begin{equation} \label{eu_eqn22}
S_{\text{D}2,T=0}^{\text{1-loop}} \sim \int  d^2x  \frac{N^2}{\beta^3 \phi^{1}} 
\end{equation}

The next step is to equate $S_{\text{D}2}^{\text{cl}}$ and  $S_{\text{D}2,T=0}^{\text{1-loop}}$. It should be noted that this in no way suggests neglecting the higher loop contributions. According to SUGRA analysis, it may mean that all loop contributions are of the same order and hence the equating is legit. It seems reasonable, since here we consider the strong-coupling region in the dual field theory. So, equating the two: 

\[ \frac{N^2}{\beta \lambda } \phi^2 = \frac{N^2}{\beta^3 \phi^{1} } \]

\[ \Rightarrow  \phi^{(3)} = \lambda \beta^{-2}  \]

\[ \phi = \lambda^\frac{1}{3} \beta^\frac{-2}{3} \]

Therefore, 

\begin{equation} \label{eu_eqn23}
  \phi = \lambda^\frac{1}{3} T^\frac{2}{3}
\end{equation}

Further, by substituting this value into $S_{\text{D}2}^{\text{cl}}  \sim S_{\text{D}2,T=0}^{\text{1-loop}} \sim S_{\text{D}2}$ , the free energy is:

\begin{equation} \label{eu_eqn24}
  F_{\text{D}2} \sim S_{\text{D}2}/\beta \sim  N^2 T^{\frac{10}{3}} \lambda^{-\frac{1}{3}} V_2
\end{equation}

where $V_2$ is the spatial volume of the D2-brane. Interestingly, the low temperature condition here is reminiscent of the small curvature condition in gravity! \cite{Itzhaki:1998ppr1}\\

This expression for free energy, which was calculated by authors of \cite{Morita:2013ppr1}, does not have a Chern-level 'k' dependence. The presence of a Romans mass (k) in the supergravity side (\cite{Jafferis:2015ppr1}), induces a Chern Simons term on the D2 brane. We would like to express the free energy expression in terms of the Chern-Simons level.

By using the 't Hooft limit $ \lambda = \frac{N}{k} $ in equation (\ref{eu_eqn24}), we get : \\

\begin{equation} \label{eu_eqn25}
F_{\text{D}2} \sim N^{\frac{5}{3}} k^{\frac{1}{3}} T^{\frac{10}{3}} V_2
\end{equation} 

Clearly, the N and k dependence of the free energy computed at finite temperature matches with what was calculated in \cite{Jafferis:2015ppr1} and \cite{Rong:2015ppr2}. \\

Our result is valid for low temperatures : $(T/ \lambda)^{\frac{1}{3}} \ll 1$ , i.e. , $(Tk/N)^{1/3} \ll 1$ . \\

We also predict that the free energy of black-brane solutions in supergravity would have a temperature dependence which can be read off from the free energy expression computed by us.

\section{Conclusions}
\label{sec:conc}
We were able to compute the expected form of free energy expression from field theory at finite temperature. The Smilga-Wiseman method helps us to obtain the correct N dependence without having to use localization techniques. It also predicts the temperature dependence of free energy in a given low energy regime, which is expected to match with results from black-brane solutions in supergravity. The method rests on certain properties of the classical moduli fields of the theory , without having to look at the entire effective theory. A better understanding of the key assumptions of the method still remains to be explored, especially why we need to consider only the one-loop potential and not the higher loop corrections. Though some good arguments have been suggested recently. it is hoped that these questions would be addressed more concretely in the near future. 
The conjectured duality in \cite{Rong:2015ppr2} was backed by the complete agreement between the free energies of their supergravity solution and the conjectured dual. They also expected this agreement to hold when the AdS vacuum is replaced by an AdS black hole, and the dual CFT is at non-zero temperature. Our field theory calculation backs this expectation further. We do hope to see black hole solutions asymptotic to their solutions some time soon, for which the free energy scales with temperature as predicted by us.
 \\

\acknowledgments

I would like to thank Pallab Basu for useful discussions and guidance over the past one year. I am also really thankful to Suvrat Raju for his constructive critique of the idea while it was still nascent. I have discussed parts of this work with various people including but not limited to Roji Pius and Sudip Ghosh, and I really appreciate their inputs. I also thank Yi Pang and Junchen Rong for online correspondence. I am particularly grateful to the hospitality at ICTS-TIFR, Bangalore from where this work ensued.

\end{document}